\documentclass[12pt]{iopart}

\usepackage{graphicx} %eps figures can be used instead
\begin{document}

\title[Frictional anisotropy of metal nanoparticles adsorbed on graphene]{Frictional anisotropy of metal nanoparticles adsorbed on graphene}

\author{A V Khomenko$^{1,2}$, N V Prodanov$^{1,3,4}$, M A Khomenko$^{1}$, B~O~Krasulya$^{1}$} 

\address{$^1$ Department of Complex Systems Modeling, Sumy State
University, 40007 Sumy, Ukraine}
\address{$^2$ Peter Gr\"unberg Institut-1, Forschungszentrum-J\"ulich, 52425 J\"ulich, Germany}
\address{$^3$ J\"ulich Supercomputing Center, Institute for Advanced Simulation, Forschungszentrum-J\"ulich, 52425 J\"ulich, Germany}
\address{$^4$ Dept. of Materials Science and Engineering, Universit\"at des Saarlandes, 66123 Saarbr\"ucken, Germany}

\ead{khom@mss.sumdu.edu.ua, prodk@rambler.ru}

\begin{abstract}
Friction force acting on silver and nickel nanoparticles sheared on a graphene sheet in different lateral directions is investigated using classical molecular dynamics simulations. The results reveal the existence of frictional anisotropy for both metals. In most cases, the maximum value of the friction force is about two times larger than the minimum one. The form of dependencies of instantaneous values of the friction force components on the corresponding lateral components of the position of the centre of mass of the nanoislands strongly depends on sliding direction, varying between the sawtooth and the irregular one. A qualitative explanation of the results based on the ``patch'' model is proposed.

\end{abstract}

\vspace{2pc}

\maketitle

\section{Introduction}\label{sec1}

The ongoing endeavour to miniaturize various devices with moving components often meets an obstacle which hampers their reliable operability and which is related to friction at the atomic level~\cite{Socoliuc2006,Gnecco2007,Bhushan2008,Pantazi2008,Pogrebnjak2009,Pogrebnjak2009vac,Guerra2010,Lyashenko2011}. A deep understanding of fundamental nanotribological processes is required in order to control friction and wear at the nanoscale and to successfully proceed with miniaturization tasks.
In this context, manipulation of metal nanoparticles (NPs) adsorbed on an atomically-flat surface using the atomic force microscope (AFM) tip is worth mentioning. This method has been developed recently for quantitative investigation of friction of antimony NPs sliding on highly oriented pyrolytic graphite (HOPG) substrate~\cite{Ritter2005,Dietz2008,Dietz2009,Schirmeisen2009,Dietz2010tl}. Compared to traditional techniques it offers several advantages. First, the method provides much wider range of materials available for consideration. Second, with AFM it is possible to probe tribological properties of nanoobjects with sizes which are not accessible to friction force microscopy (FFM) and surface force apparatus. The manipulation experiments reveal the dual behaviour of the friction force $F_{\mathrm{f}}$ acting on Sb NPs under ultra-high vacuum conditions. Some NPs exhibit vanishing friction while others experience finite $F_{\mathrm{f}}$ which depends linearly on the contact area $A$~\cite{Dietz2010}. Note that in the experiments typical diameter of a NP varies from about 50~nm up to a few 100~nm, and the contact area is deduced from lateral dimensions of the NP so the flat contact interface is assumed.

From the theoretical side, several works pertaining to friction of metal NPs on a solid substrate can be found in literature~\cite{Guerra2010,Aruliah2005,Brndiar2011,Khomenko2010jpc,Khomenko2013}. Seminumerical model~\cite{Aruliah2005} is aimed at calculating the power necessary to depin a NP on a surface with an AFM tip in tapping mode. However, a large amount of assumptions, such as, two-dimensionality of the system, the rigidity of the NP and the substrate, quasistatic motion and the absence of the $F_{\mathrm{f}}$ measurements make it inappropriate for comparisons to the kinetic friction experiments. Molecular dynamics (MD) simulations of the ballistic friction~\cite{Guerra2010} consider sliding of a golden NP on the HOPG substrate with high velocities and do not measure $F_{\mathrm{f}}$. Therefore, this model cannot describe the manipulation experiments. \textit{Ab initio} calculations of friction of a Sb$_4$ molecule on HOPG surface~\cite{Brndiar2011} make an attempt to explain the dual behaviour of $F_{\mathrm{f}}$ in the experiments. However, it is questionable whether the results for one Sb molecule can be transferred to the relatively large NPs comprising hundreds of thousands and more atoms. In addition, sliding in a static manner~\cite{Brndiar2011} makes it questionable whether it is valid to compare the calculations~\cite{Brndiar2011} with the experiments, where NPs move with velocities of the order of $\rm \mu m/s$.

MD simulations of friction of silver and nickel~\cite{Khomenko2010jpc} as well as golden and copper~\cite{Khomenko2013} nanoislands adsorbed on a graphene sheet provide a model which is relatively close to the manipulation experiments. Although the typical lateral dimension of the largest NP is about 10~nm which is 5 times smaller than in the quantitative tribological experiments, such NPs are used in other manipulation trials~\cite{Baur1998,Paolicelli2009,Rovatti2011}. Further enhancement of sensitivity of the equipment will give the opportunity to perform quantitative measurements for such nanoislands and to compare with computational results. Another deviation of the model~\cite{Khomenko2010jpc} from the experiments is the use of different materials, viz graphene as the substrate and Ag or Ni for the NPs. Nevertheless, such an approach should reproduce the general behaviour of the experimental systems, because their features which are essential to tribological properties should not differ significantly from our model. For example, the structural contribution of graphene to friction of nanoislands should be close to that of HOPG because these materials have identical lattices. Friction and other tribological properties of graphene itself have attracted significant attention in the last years~\cite{Filleter2009,Lee2009,Lee2010,Wijn2011,JNEPH2009} and our simulations may provide valuable information about it. Additionally, the development of new methods of production of Ag~\cite{Abargues2009,Jeon2010}, Ni~\cite{Geissler2010} and Pt~\cite{Nethravathi2011} NPs with sizes comparable to our system provide the possibility for experiments which directly correspond to our model.

The main aim of the simulations in Refs~\cite{Khomenko2010jpc,Khomenko2013} was to define the dependence of the friction force acting on NPs on contact area. However, many other unresolved questions have remained. One of them is the dependence of the friction of nanoislands on the sliding direction relatively to the honeycomb lattice of graphene. The existence of the frictional anisotropy in different systems has attracted a significant attention recently~\cite{Park2005,Filippov2010,Gnecco2010,Braun2011,Marcus2002} because it allows one to separate the dominating contributions to friction from a number of other factors. It can also help clarify the origin of the superlubricity phenomenon~\cite{Gnecco2010,Khomenko2010carbon}. According to the experiments~\cite{Park2005} and simulations~\cite{Filippov2010,Gnecco2010,Braun2011}, frictional anisotropy usually exists in small systems where the contribution of the atomic structure of the interface cannot be neglected. However, it is not necessary for both contacting surfaces to have periodic crystal lattice in order to observe the phenomenon as was shown for quasicrystals~\cite{Park2005,Filippov2010} and lubricated systems~\cite{Braun2011}. One should note that the results mentioned above were obtained for kinetic friction, while static friction in the one-atom Prandtl-Tomlinson model for the hexagonal lattice of the substrate is isotropic~\cite{Gnecco2010}.

According to the manipulation experiments, Sb nanoislands do not manifest frictional anisotropy on HOPG surface~\cite{Dietz2010}. Additionally, no locked state of NPs was found during their rotation. These results may be caused by the following two factors. First, relatively large sizes of nanoislands with hundreds of thousands of atoms in contact may lead to washing out of the anisotropy effect. Second, the nearest-neighbour distance in Sb is much larger than in graphene and may prevent the interface from forming locked regions. In order to clarify these questions, we have carried out MD simulations described in the present paper, where the direction dependence of the kinetic friction of silver or nickel nanoisland sliding on graphene is investigated.

\section{Model}\label{model}

\begin{figure}[htb]
\centerline{\includegraphics[width=0.35\textwidth]{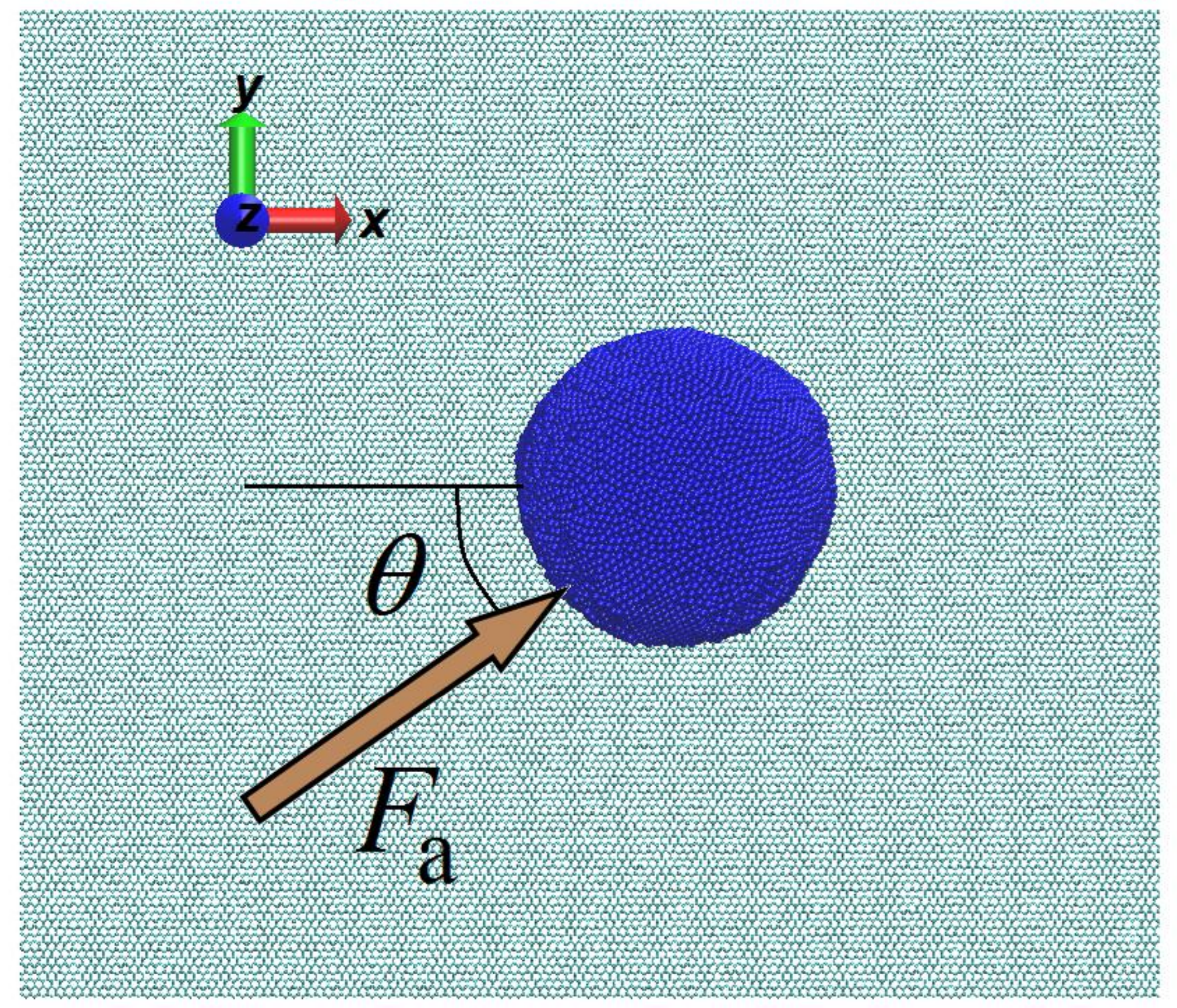}
\includegraphics[width=0.28\textwidth]{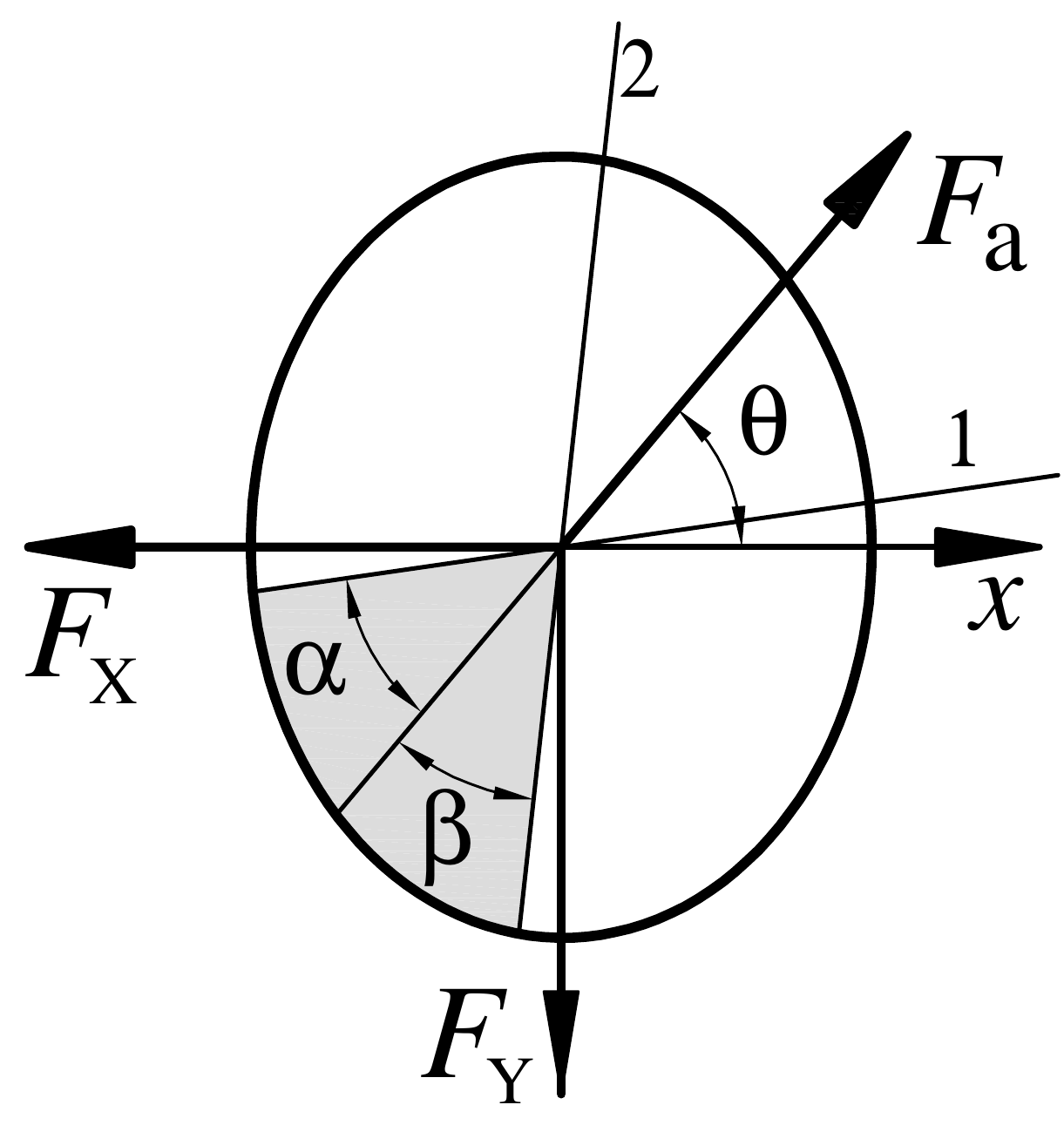}}
\caption{Top view of the system with a Ni nanoisland containing 29000 atoms (left). Schematic sketch of the forces acting on the NP during its shear (right). (All snapshots in this work were produced with VMD software~\cite{Humphrey1996}).}
\label{snapshots}
\end{figure}

The details of the model can be found elsewhere~\cite{Khomenko2010jpc,JNEPH2011}. Here we point out only its main peculiarities. A graphene sample has zigzag and armchair edges parallel to the \textit{x}- and \textit{y}-directions, respectively (see figure~\ref{snapshots}). To maintain the constant position of the graphene in the $z$-direction, boundary carbon atoms along the perimeter of the layer are kept rigid in the simulations. Ni and Ag NPs containing 10000, 20000, 25000 and 29000 atoms are considered. For each size of the NP the unique $x \times y$ dimensions of the substrate are used, varying from 20.7$\times$17.9~nm to 37.4$\times$32.4~nm. The total number of atoms in the simulations is in the range from 24112 to 75208.

Graphene atoms interact with each other via the harmonic potential~\cite{Sasaki1996}. Forces between metal atoms are derived from the alloy form of the embedded atom method potential~\cite{Zhou2001}. The pairwise 6-12 Lennard-Jones potential with the parameters from~\cite{Khomenko2010jpc} describes the metal-carbon interactions. The equations of motion are integrated using the leapfrog method~\cite{Rapaport2004,Griebel2007} with a time step $\Delta t = 0.2$~fs. Calculations were performed on GPUs NVIDIA GeForce GTX 460 and GeForce GTX 480.

\begin{figure}[htb]
\centerline{\includegraphics[width=0.55\textwidth]{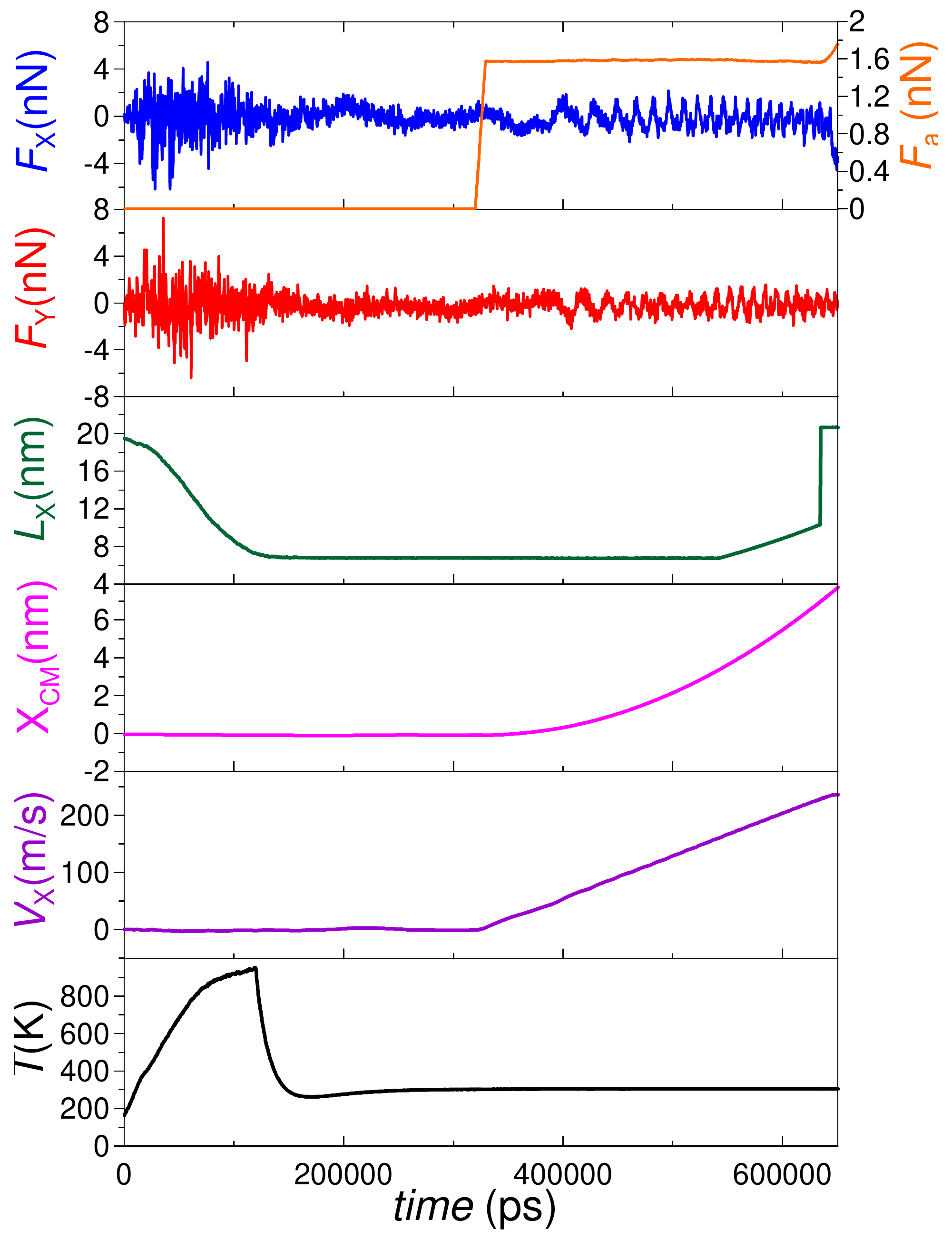}}
\caption{Time dependencies of quantities measured in the simulation of Ni NP containing 10000 atoms sheared in the direction corresponding to $\theta=30^{\circ}$. Here, $F_{\mathrm{X}}$, $F_{\mathrm{Y}}$ -- lateral components of the substrate force; $L_{\mathrm{X}}$ -- lateral size of the NP in the $x$-direction; $X_{\mathrm{CM}}$, $V_{\mathrm{X}}$ -- $x$-component of the position and the velocity of the centre of mass of the NP, respectively; $T$ -- temperature of the system. In the topmost plot the horizontal line is the applied force $F_{\mathrm{a}}$ and the second (irregular) one is $F_{\mathrm{X}}$.}
\label{time_dep}
\end{figure}

A general flow of a typical simulation can be deduced from the figure~\ref{time_dep}. A nanoisland is obtained by the procedure imitating the dewetting of thin metallic films by thermal treatments~\cite{Khomenko2010jpc,Geissler2010,Neek2009}. In this approach, a thin slab of metal atoms packed into an ideal fcc lattice is placed initially on the graphene. After the start of the calculations, it self-organizes into more compact nanoisland through the exothermal process. When the lateral sizes $L_{\mathrm{X}}$ and $L_{\mathrm{Y}}$ of the NP diminish enough and the NP gains a desired symmetric elliptical shape, we apply the Berendsen thermostat~\cite{Griebel2007} both to metal and the graphene during a suitable time interval to cool the system down to the temperature of about 300~K. After that the thermostat is decoupled from the metal atoms and is applied only to the graphene to dissipate the heat generated during the shear of the NP. The sizes $L_{\mathrm{X}}$ and $L_{\mathrm{Y}}$ are computed as the difference between the coordinates of metal atoms with maximum and minimum values along the corresponding direction. Values of $L_{\mathrm{X}} \times L_{\mathrm{Y}}$ are about 6.7$\times$6.3, 7.6$\times$8.0, 8.4$\times$8.5, 8.9$\times$8.9~nm for Ni NPs containing 10000, 20000, 25000, 29000 atoms, respectively. For corresponding Ag NPs these quantities are 7.1$\times$7.2, 8.6$\times$9.1, 9.0$\times$10.0, 10.0$\times$10.2~nm, respectively.

After the nanoisland has formed, it is sheared by the force $\mathbf{F}_{\mathrm{a}}$ imitating pushing with the AFM tip in the manipulation experiments. Shear force is applied to every atom located in the hatched sector in the right figure~\ref{snapshots} along the specified direction defined by the angle $\theta$ (counted from the \textit{x} direction). The sector is formed by the lines 1 and 2, and the angles $\alpha = \beta = 30^{\circ}$ provide approximately the same number of atoms in each half of the sector (symmetrically to the shear direction). Such a geometry gives zero torque relative to the NPs centre of mass (CM) in the \textit{xy}-plane, thus preventing the NP from rotating in this plane. A metal atom belongs to the hatched area if its coordinates satisfy the conditions following from the equations of the lines 1 and 2:

\begin{equation}\label{lines}
    \left\{
    \begin{array}{c}
      y \leq x \tan\left(\theta - \alpha \right),\\
      y \geq x \tan\left(\theta + \beta \right).
        \end{array}\right.
\end{equation}

At the beginning of a simulation run, \textit{x}- and \textit{y}-components of the applied force are incremented in steps of 0.0001 pN until the \textit{x}-component of the velocity of the CM $V_{\mathrm{X}}$ of the NP reaches the value of 3.55~m/s. Then the applied force acting on each atom in the sector remains constant, and the simulations are held with constant total applied force $\mathbf{F}_{\mathrm{a}}$. In the computer experiments both lateral components $F_{\mathrm{X}}$ and $F_{\mathrm{Y}}$ of the substrate force and the coordinates of the CM of the NP $X_{\mathrm{CM}}$, $Y_{\mathrm{CM}}$ are measured. Each force component is defined as the sum of the corresponding components of forces acting on metal atoms from the graphene. The total friction force $F_{\mathrm{f}} = \sqrt{\langle F_{\mathrm{X}}\rangle^2 + \langle F_{\mathrm{Y}}\rangle^2}$ is defined by the time-averaged values of the lateral components of the substrate force. We consider the following values of $\theta$: 0$^{\circ}$, 22.5$^{\circ}$, 30$^{\circ}$, 45$^{\circ}$, 60$^{\circ}$, 67.5$^{\circ}$ and (for some NPs) 89$^{\circ}$. One should note that MD simulations do not allow one to investigate the low NP sliding velocities (of order $\sim {\rm \mu m/s}$) used in the AFM experiments. However, the simulation results obtained in Refs.~\cite{Khomenko2013,nap2012} suggest a very weak dependence on velocity in the studied range.

\section{Results}\label{results}

\begin{figure}[htb]
\centerline{\includegraphics[width=0.45\textwidth]{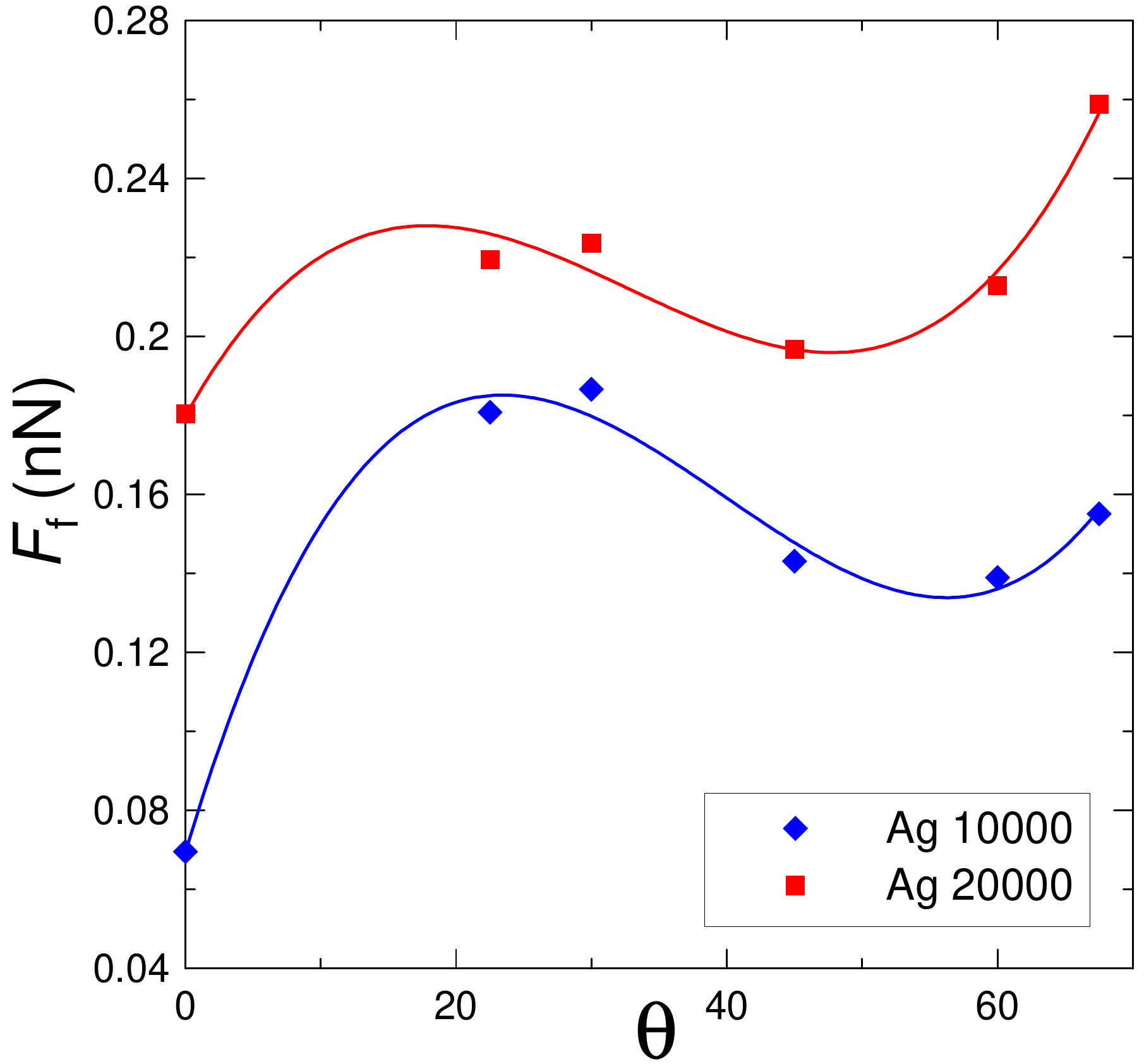}
\includegraphics[width=0.45\textwidth]{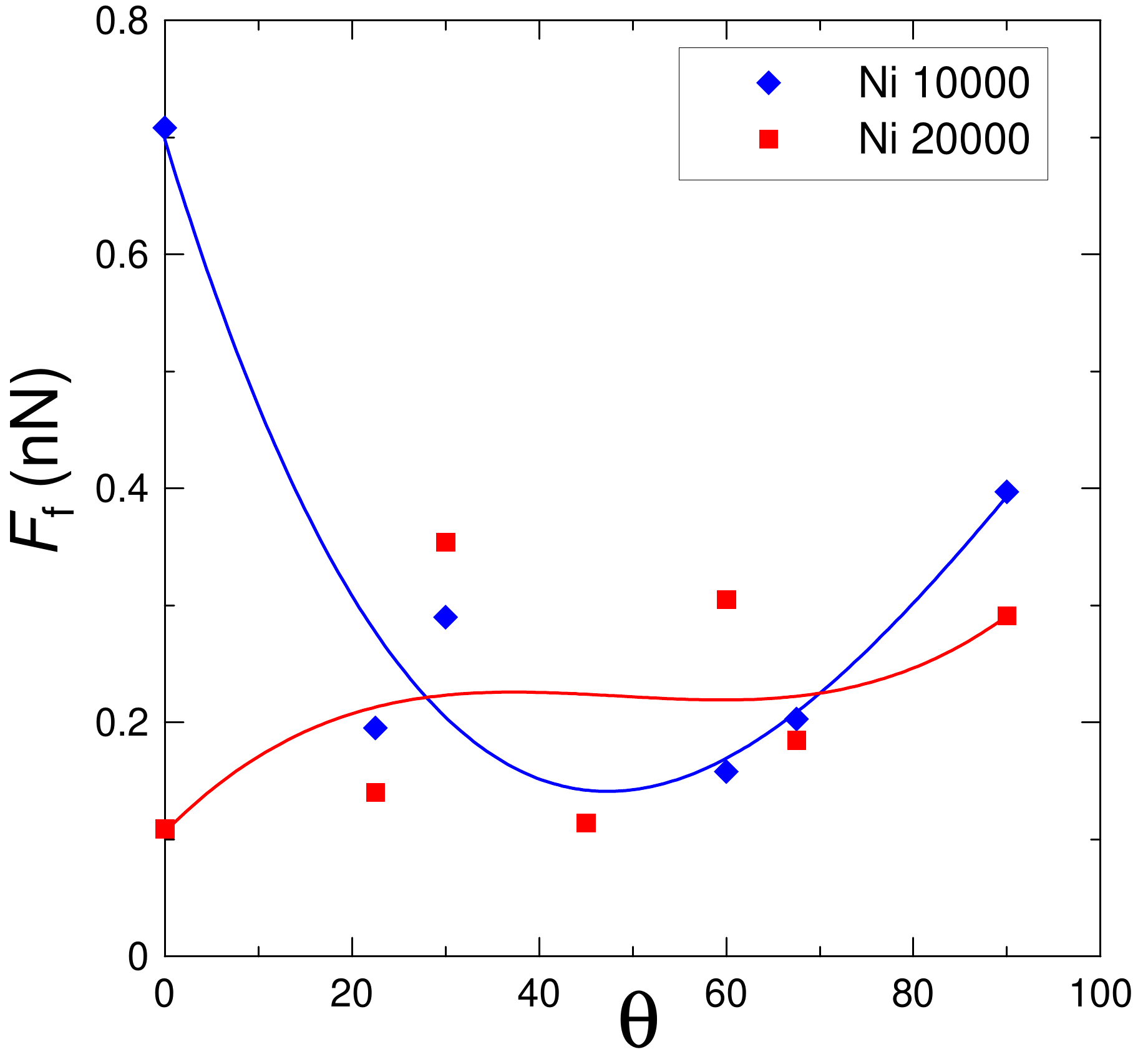}}
\caption{The total friction force acting on Ag and Ni nanoislands containing 10000 and 20000 atoms for different sliding directions.}
\label{F_theta_10000}
\end{figure}

\begin{figure}[htb]
\centerline{\includegraphics[width=0.45\textwidth]{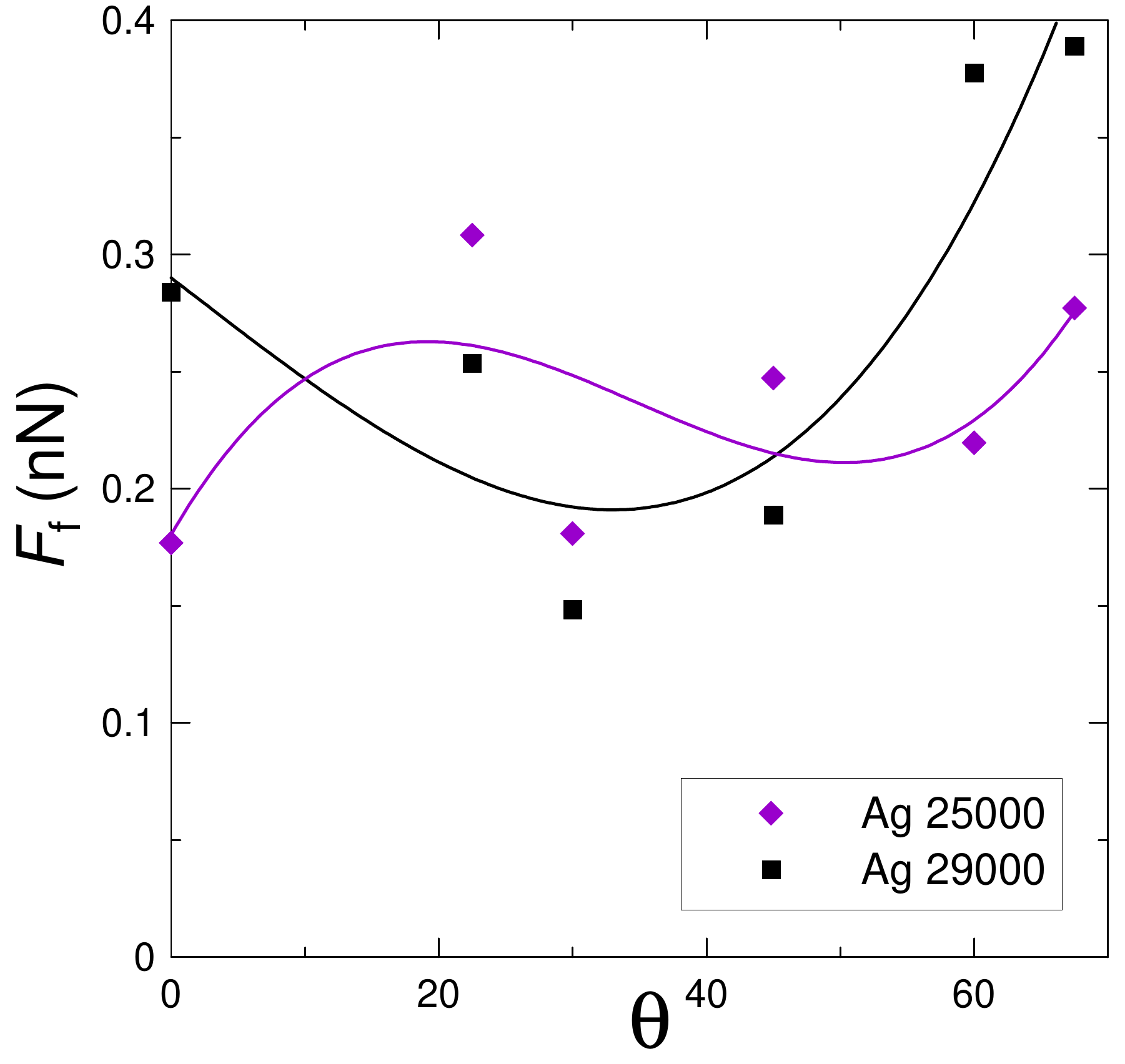}
\includegraphics[width=0.45\textwidth]{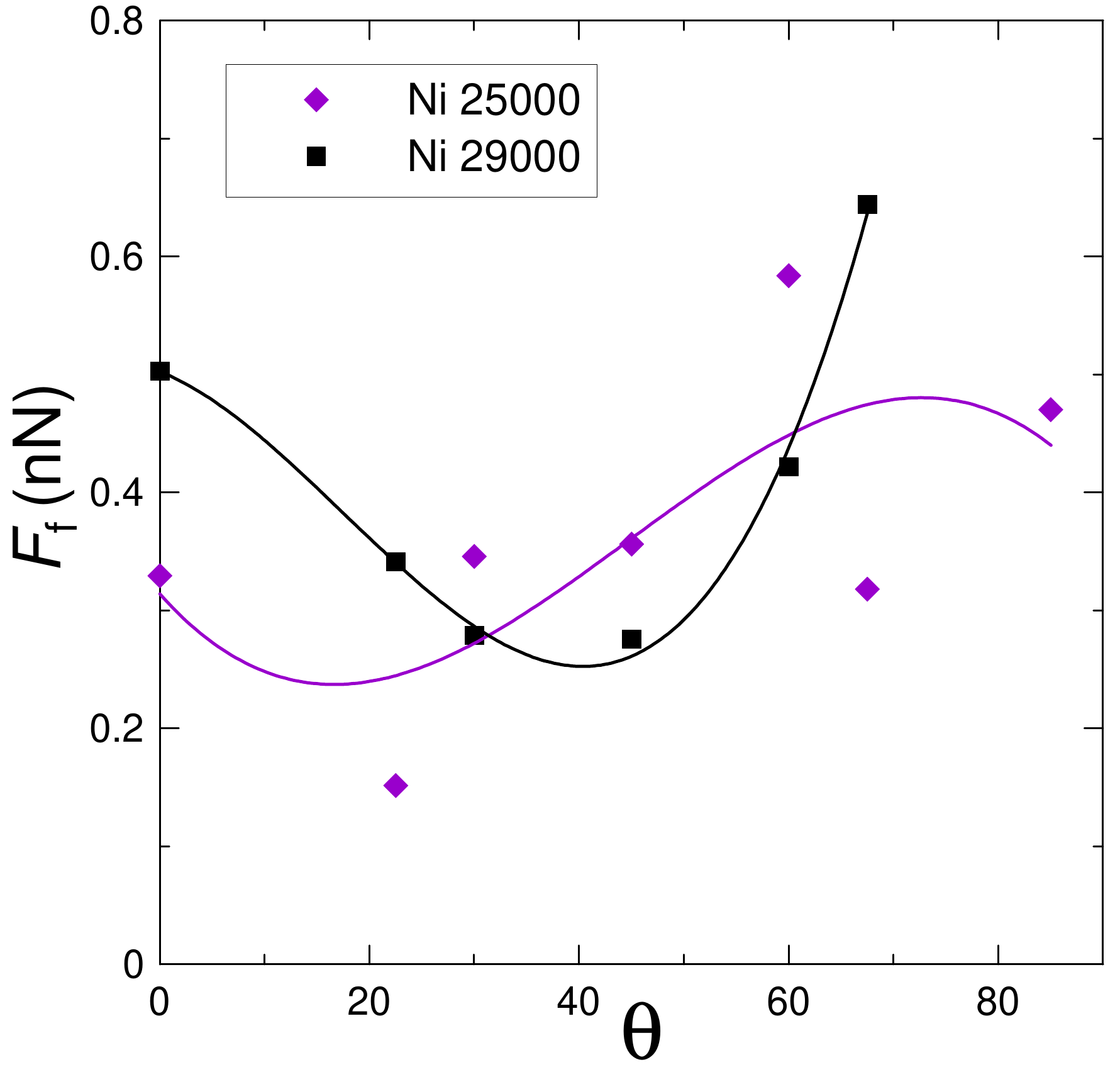}}
\caption{Direction dependencies of the total friction force acting on nanoislands containing 25000 and 29000 atoms.}
\label{F_theta_25000}
\end{figure}

\begin{figure}[htb]
\centerline{\includegraphics[width=0.48\textwidth]{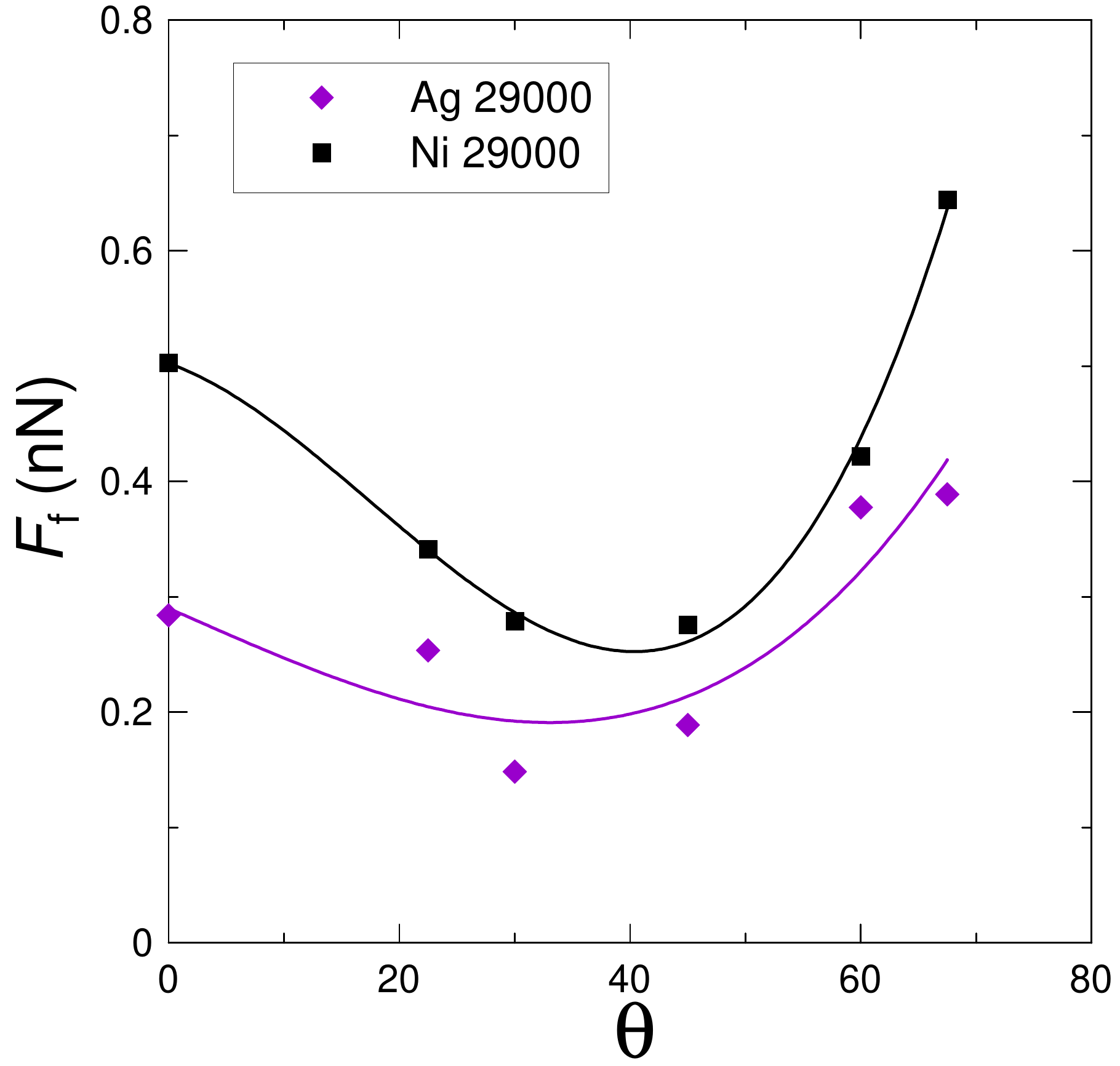}}
\caption{Total friction force vs angle $\theta$ for Ag and Ni NPs containing 29000 atoms.}
\label{F_theta_29000}
\end{figure}

Let us analyze the dependencies of the total friction force $F_{\mathrm{f}}$ on the angle $\theta$ displayed in figures~\ref{F_theta_10000}, \ref{F_theta_25000}, \ref{F_theta_29000}. Markers in the figures correspond to the calculated values and the solid lines are cubic splines shown to follow the eye. Generally, more or less pronounced changes of $F_{\mathrm{f}}$ with $\theta$ are observed for all considered NPs. For most cases, the maximum value of the friction force is about two times larger than the minimum one. There is a correlation in variations of $F_{\mathrm{f}}$ with $\theta$ for Ag NPs containing 10000 and 20000 atoms (figure~\ref{F_theta_10000}, left part), with local maxima in the angle range between about 18$^{\circ}$ and 30$^{\circ}$. However, for the corresponding Ni NPs the picture is different with considerable data scatter (cf. figure~\ref{F_theta_10000}, right part). Such a result can be attributed to the fact that the nearest-neighbor distance $d_{\mathrm{Ni}}=0.249$~nm~in Ni is very close to the graphene lattice constant $a = 0.246$~nm. This may lead to a high sensitivity of $F_{\mathrm{f}}$ to small variations of number of metal atoms in contact as there is a high probability of the formation of locally commensurable interface leading to high friction~\cite{Khomenko2010jpc,Khomenko2013}. The procedure used in our model for production of nanoislands did not allow us to precisely control the shape and the area of contact of NPs, and the random factor is always present. So NPs containing the same number of atoms do not have the identical contact morphologies in different simulation runs. This fact for small Ni NPs may lead to considerable deviations in the friction force.

Data shown in figure~\ref{F_theta_25000} also indicates the existence of variations of $F_{\mathrm{f}}$ with the direction of sliding for larger NPs. The maximum and the minimum values of $F_{\mathrm{f}}$ differ by about factor of 2. This value is typical for many systems, but it is smaller than the one found for quasicrystal substrate, where it is equal to 8~\cite{Park2005}. Data scatter did not allow us to certainly extract the empirical dependence $F_{\mathrm{f}}(\theta)$. Considerable variations of $F_{\mathrm{f}}$ with misfit angles of substrates were also observed in the lubricated systems~\cite{Braun2011} especially for the thinnest, one atom thick lubricant film. In this case, the origin of variations is the commensurability of the lubricant-substrate interface due to solidlike or liquidlike state of the film. In our model the reason of variations of friction force is different, as the contact interface of the NP does not have long-range ordering of atoms (see below). Data scatter may be ascribed to slightly different contact morphologies for NPs containing the same number of atoms in different simulation runs, which may lead to random occurrences of uncontrollable locally commensurable contacts. For Ag and Ni nanoislands composed of 29000 atoms qualitatively similar trend of $F_{\mathrm{f}}(\theta)$ dependence can be observed (figure~\ref{F_theta_29000}) with the minimum between 30$^{\circ}$ and 45$^{\circ}$.

\begin{figure}[htb]
\centerline{
\includegraphics[width=0.45\textwidth]{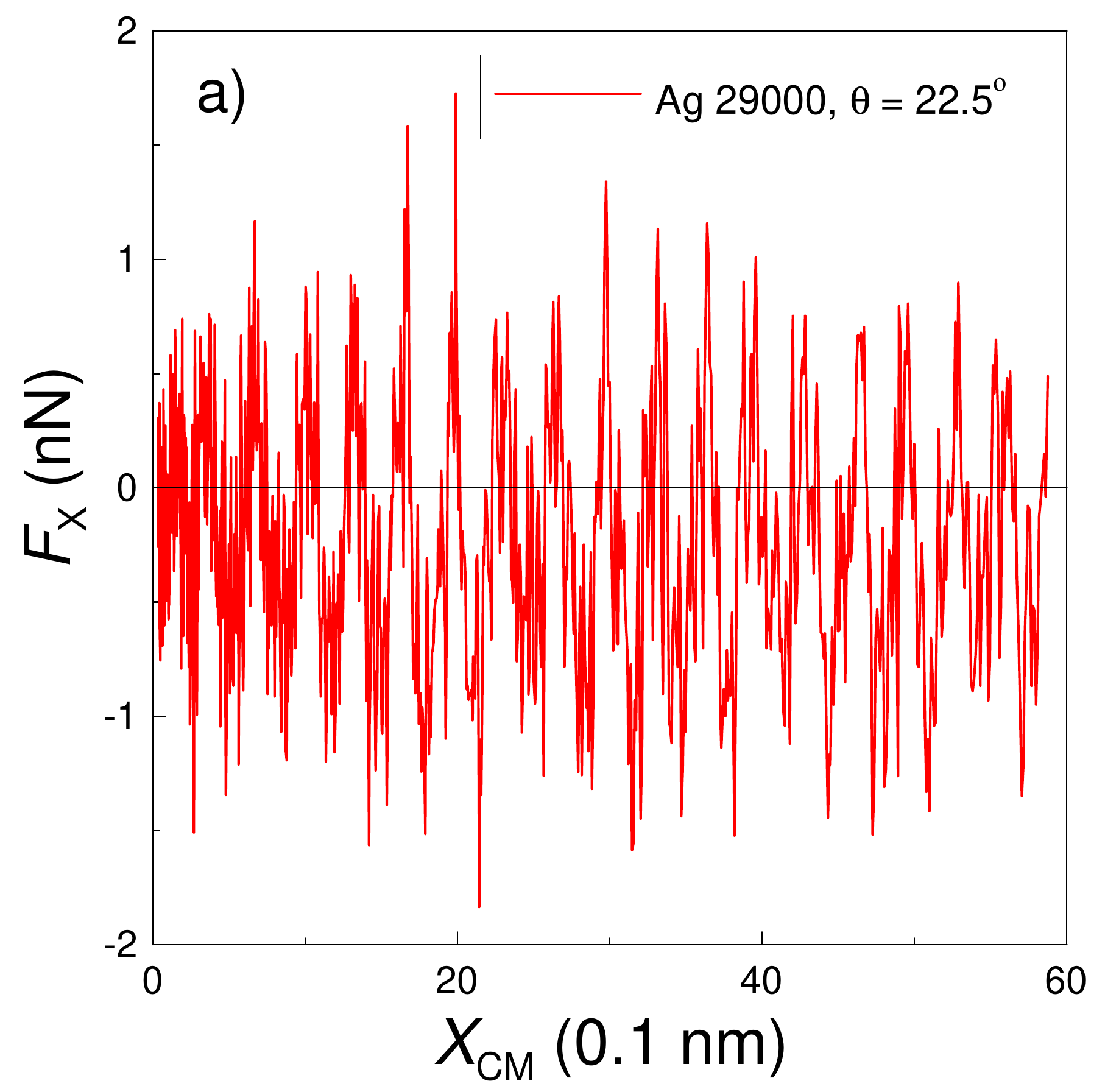}
\includegraphics[width=0.45\textwidth]{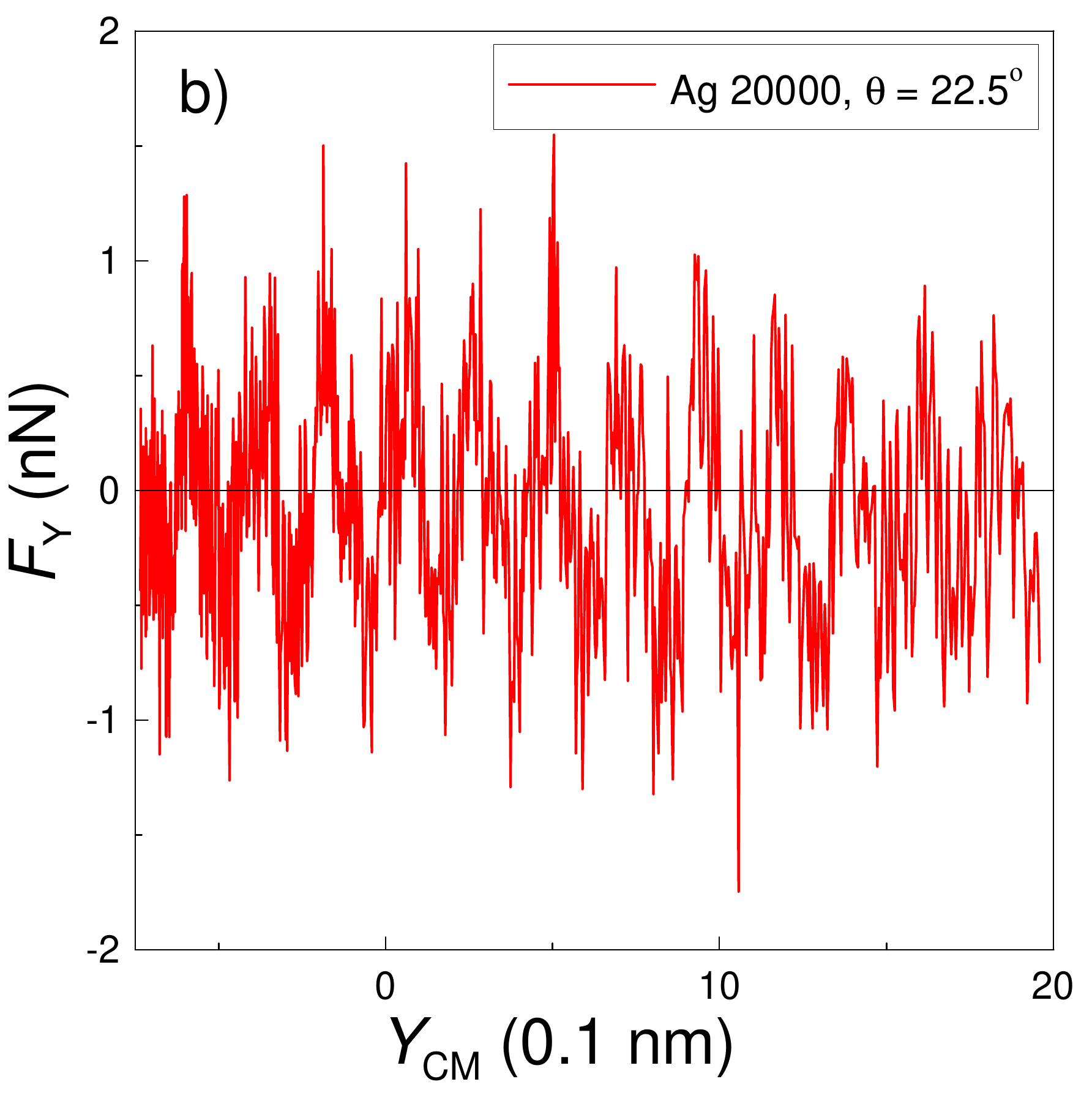}}
\centerline{
\includegraphics[width=0.45\textwidth]{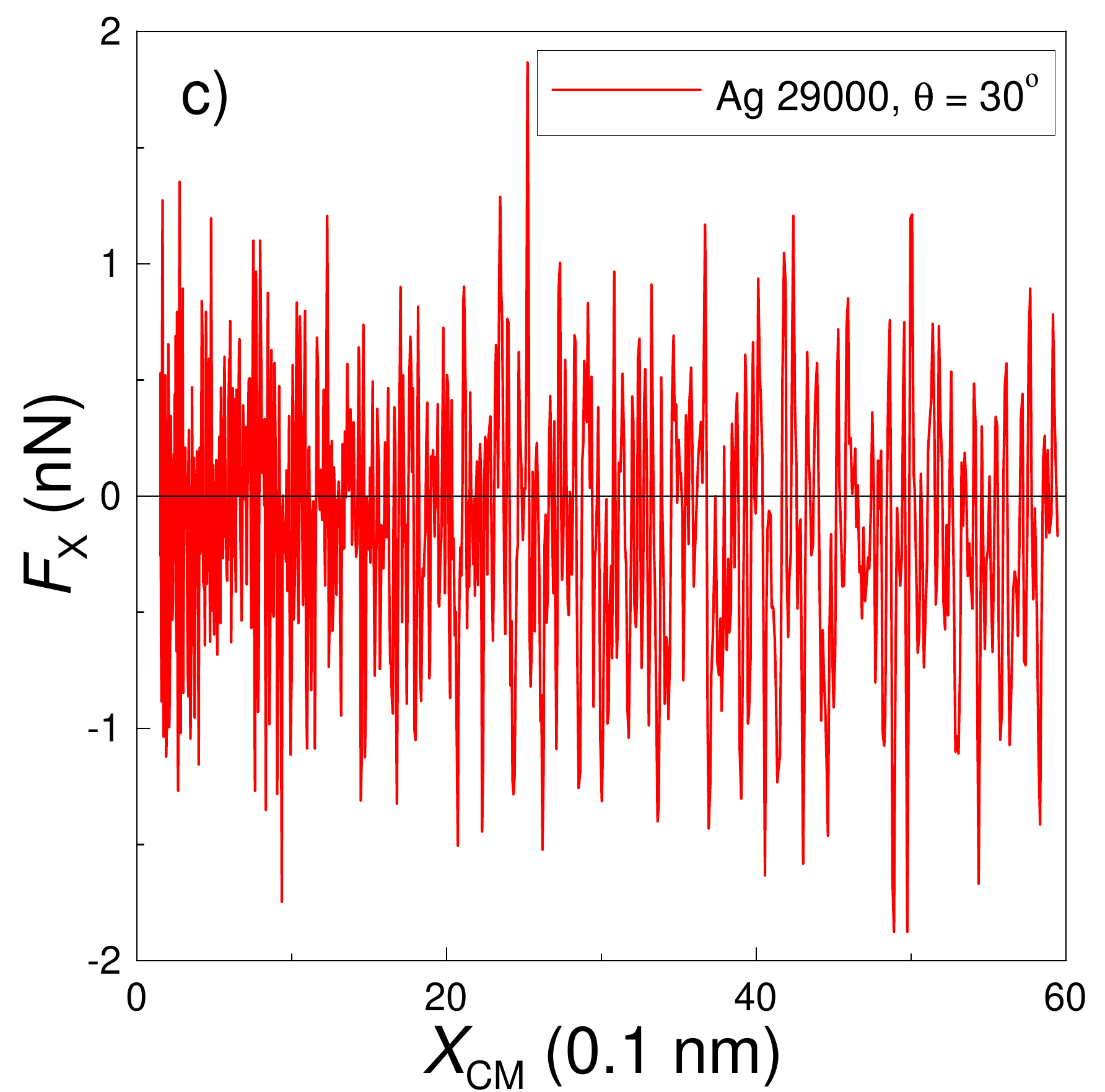}
\includegraphics[width=0.45\textwidth]{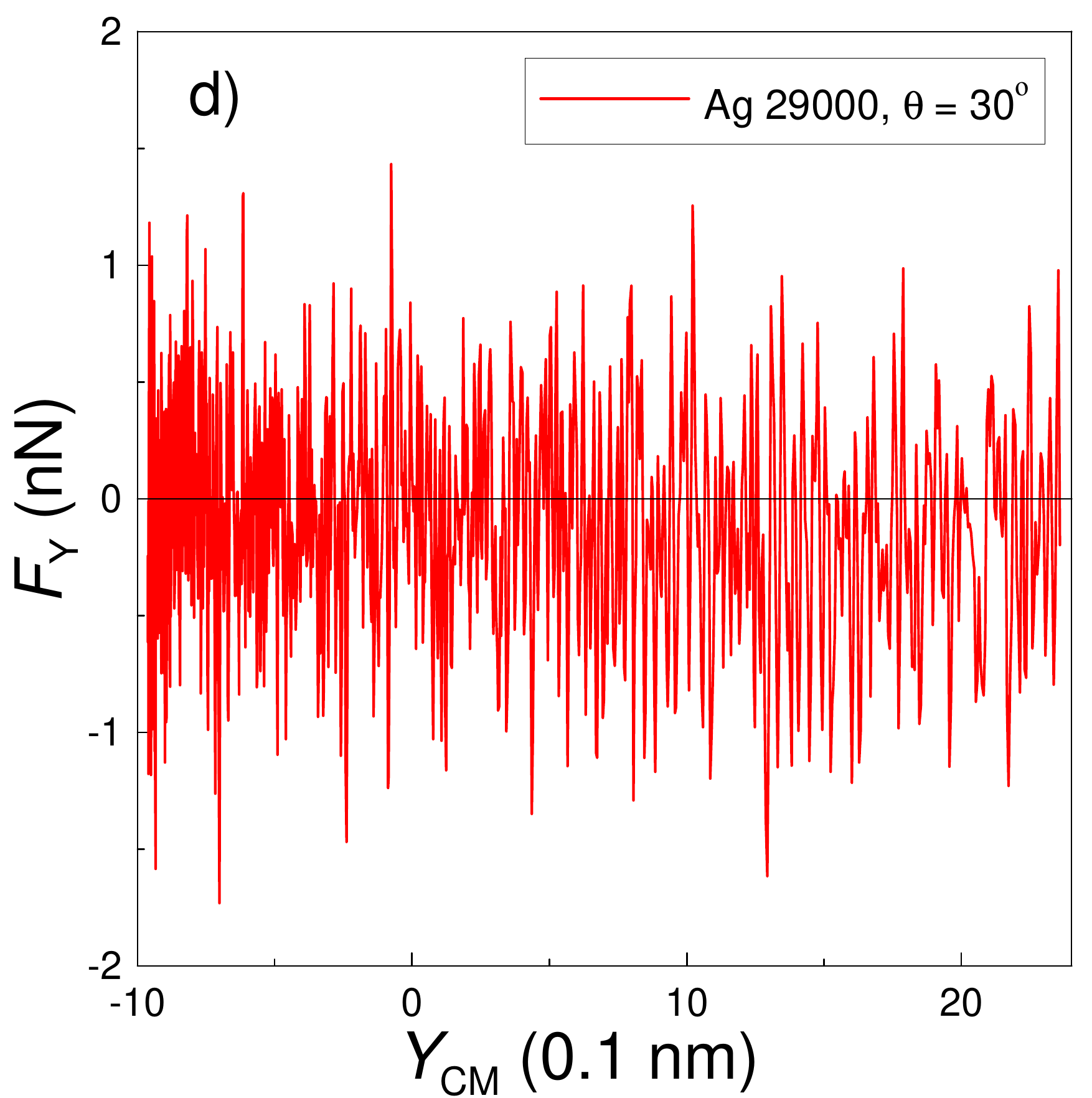}}
\caption{Examples of sawtooth (a, b) and irregular (c, d) dependencies of the substrate force components for Ag NPs.}
\label{F_Ag}
\end{figure}

\begin{figure}[htb]
\centerline{\includegraphics[width=0.45\textwidth]{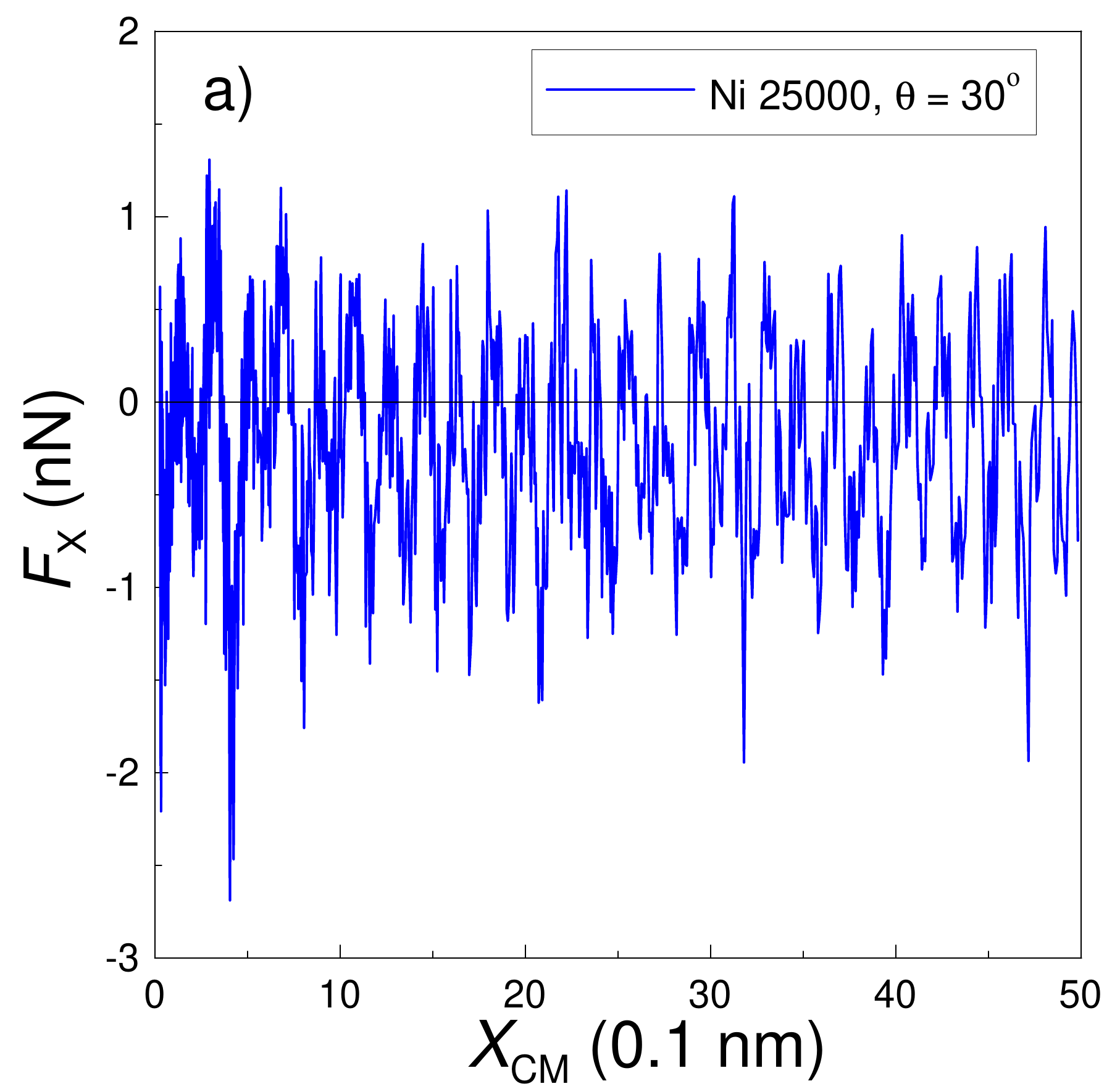}
\includegraphics[width=0.45\textwidth]{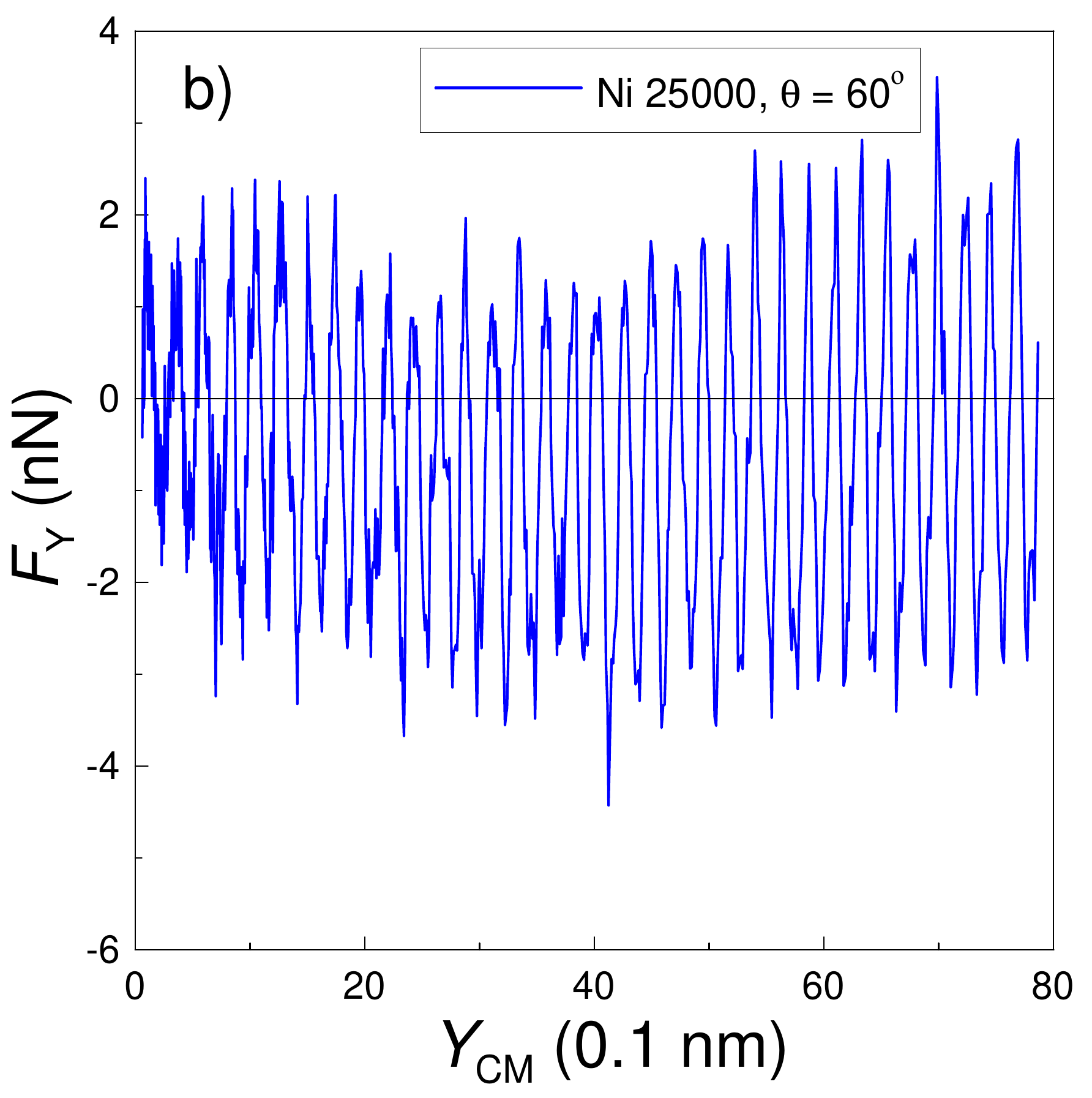}}
\centerline{\includegraphics[width=0.45\textwidth]{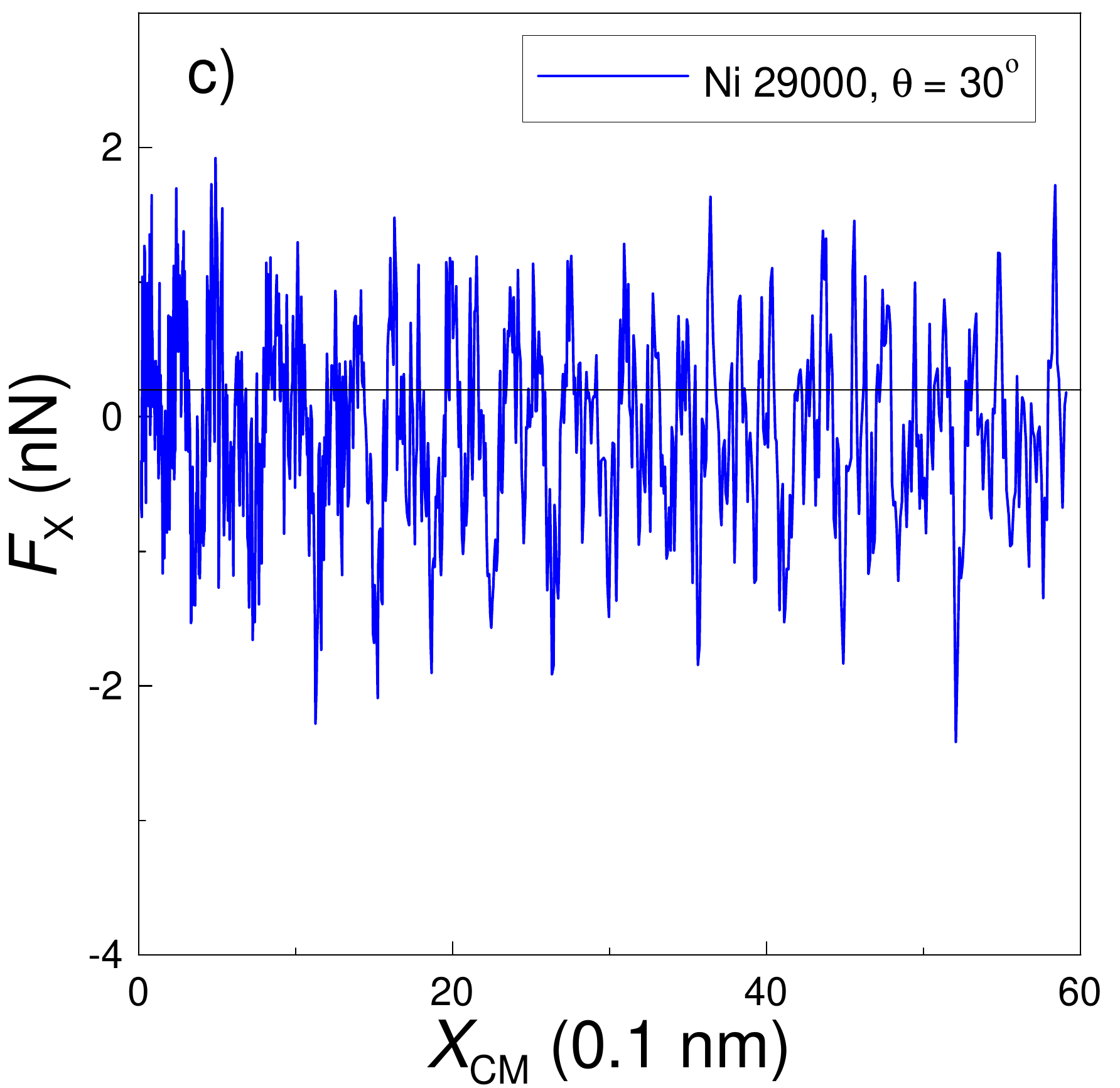}
\includegraphics[width=0.45\textwidth]{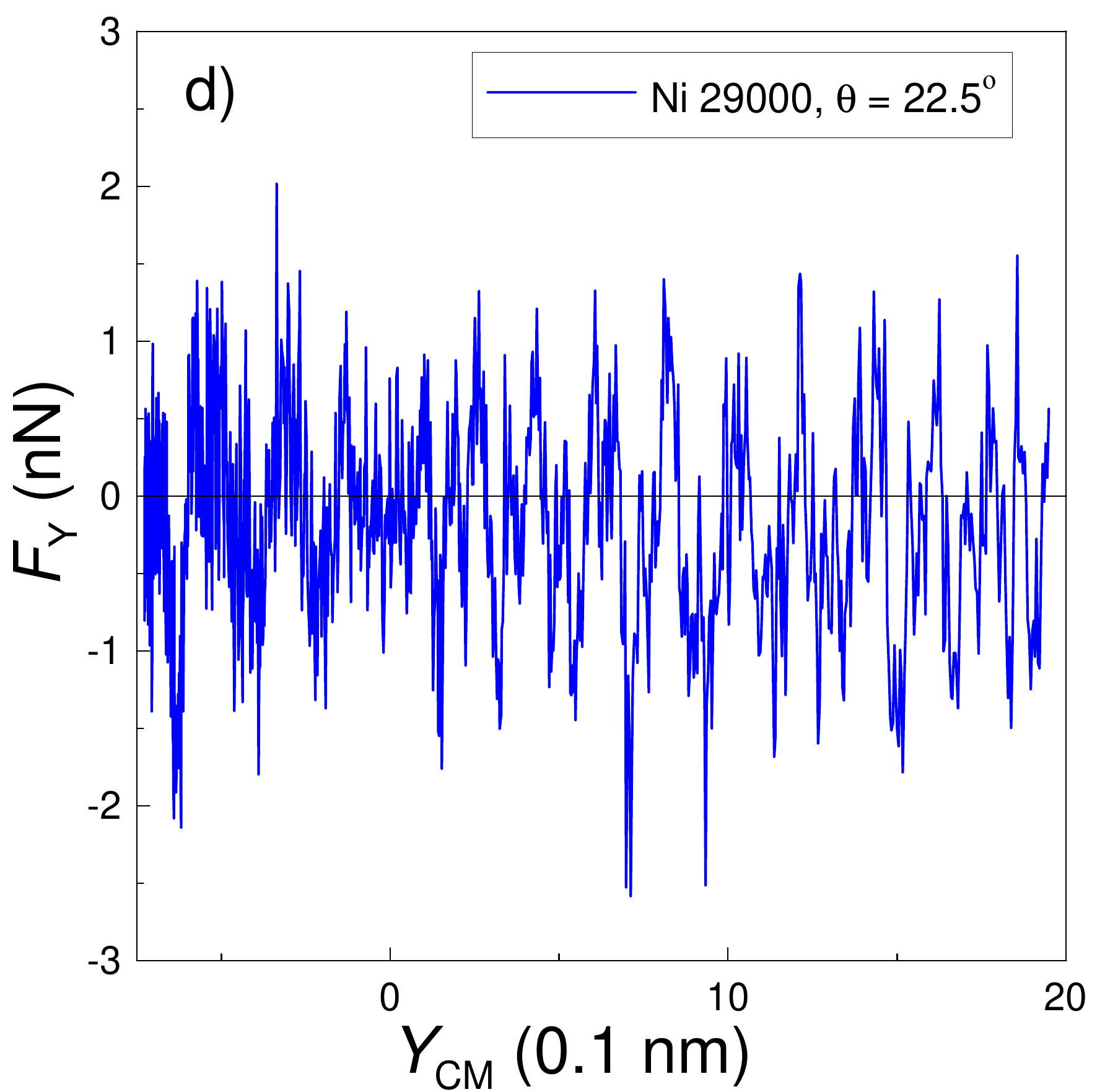}}
\caption{Sawtooth (a, b) and irregular (c, d) dependencies of the substrate force components for Ni NPs.}
\label{F_Ni}
\end{figure}

Let us analyze the dependencies of the lateral (substrate) force components $F_{\mathrm{X}}$ and $F_{\mathrm{Y}}$ on the corresponding components $X_{\mathrm{CM}}$, $Y_{\mathrm{CM}}$ of the CM of NPs shown in figure~\ref{F_Ag} and figure~\ref{F_Ni}. In the previous study~\cite{Khomenko2010jpc} it was found that $F_{\mathrm{X}}(X_{\mathrm{CM}})$ has an irregular shape for Ag nanoislands sheared along the direction corresponding to $\theta = 0$, while this dependence is sawtooth or has sawtooth regions for most Ni NPs. In the present work we show that both $F_{\mathrm{X}}$ and $F_{\mathrm{Y}}$ components can be sawtooth or have such regions for both metals when $\theta \neq 0^{\circ}$. This can already be seen from the time dependencies in figure~\ref{time_dep}, where both $F_{\mathrm{X}}$ and $F_{\mathrm{Y}}$ are sawtooth for Ni nanoisland containing 10000 atoms and $\theta = 30^{\circ}$. Most dependencies are sawtooth or have such regions for Ni NPs, while for Ag such a dependence is an exception rather than a rule.

As can be seen from figure~\ref{F_Ag}\textit{a},\textit{b}, shearing Ag NPs with 29000 or 20000 atoms in the direction corresponding to $\theta = 22.5^{\circ}$ leads to sawtooth components of the friction force, while other components (not shown here) are not clearly sawtooth. The distance between spikes fluctuates around the value of about 0.5~nm. Previously~\cite{Khomenko2010jpc} it was shown that the period of spikes for Ni was close to the value of the lattice constant $a$ of graphene. Deviation of the shear direction of Ag nanoisland with 29000 atoms from $\theta = 22.5^{\circ}$ to $\theta = 30^{\circ}$ leads to the qualitative change of the dependencies (figure~\ref{F_Ag}\textit{c},\textit{d}) from sawtooth to the irregular one.

In the case of Ni nanoislands the dependence of at least one of the components of the friction force is sawtooth or has sawtooth regions for the majority of calculations. In some simulation runs both components are sawtooth as is the case for Ni NPs with 10000 and 25000 and $\theta = 30^{\circ}$, while only one component is not irregular in others. In general, the period of spikes differs from the Ag NPs having the value of about 0.25~nm~close to $a$~\cite{Khomenko2010jpc}.

Our results clearly indicate that the friction force changes with the sliding direction. To explain this fact, several factors can be considered. The first one is the structure of the contact interface. We calculated radial distribution functions (RDFs) to look at the structure of nanoislands. For all NPs RDFs are analogous to the obtained previously~\cite{Khomenko2010jpc,Khomenko2013} (they are not shown here) and have the first spike indicating the short-range order in NPs. The location of the spike is close to the nearest-neighbor distance 0.249~nm~and 0.288~nm~for Ni and Ag, respectively. Farther spikes are smeared, indicating the absence of the long-range order and suggesting that the bulk of NPs is either amorphous or polycrystalline. Visual inspection of the contact surface of nanoislands confirms this conclusion and suggests that the NP interface contains domains with crystal structure and some disordered regions.

The second factor is the influence of the energy dissipation mechanisms. According to the experiments~\cite{Park2005}, long-range order of both surfaces is not required to observe the frictional anisotropy, and it can occur when one of the sliding surfaces is amorphous. This was shown for TiN AFM tip passivated with a molecular layer of hexadecane thiol and sliding over the Al-Ni-Co single quasicrystal surface. The authors found that the friction along the aperiodic direction of the surface is eight times smaller than the force along the periodic direction. They explain such a behaviour by energy dissipation through electronic and phononic mechanisms, which are known to be highly anisotropic in the considered quasicrystals. The authors rule out the structural contribution due to an anisotropic response of the hexadecane thiol molecules that coat the tip, because the difference between vertical corrugations along periodic and aperiodic directions is small compared to the size of the alkyl chains. They also state that incommensurability between the probe and the surface is unprobable in any scan direction because the TiN tip is amorphous and is covered by thiol molecules. In contrast to the reasoning presented above, the simulations suggest that the commensurability factor can play an important role in similar systems~\cite{Filippov2010,Pers_2006}.

In our case the mentioned factors can also be considered in order to explain the obtained results. However, unlike the Ref.~\cite{Park2005}, the phononic and electronic contributions cannot lead to anisotropic effects because in our model they are taken into account through the Berendsen thermostat. The latter acts on all carbon atoms, so the energy is dissipated isotropically. Thus, the structural contribution is the most significant one in our model.

\begin{figure}[htb]
\centerline{\includegraphics[width=0.38\textwidth]{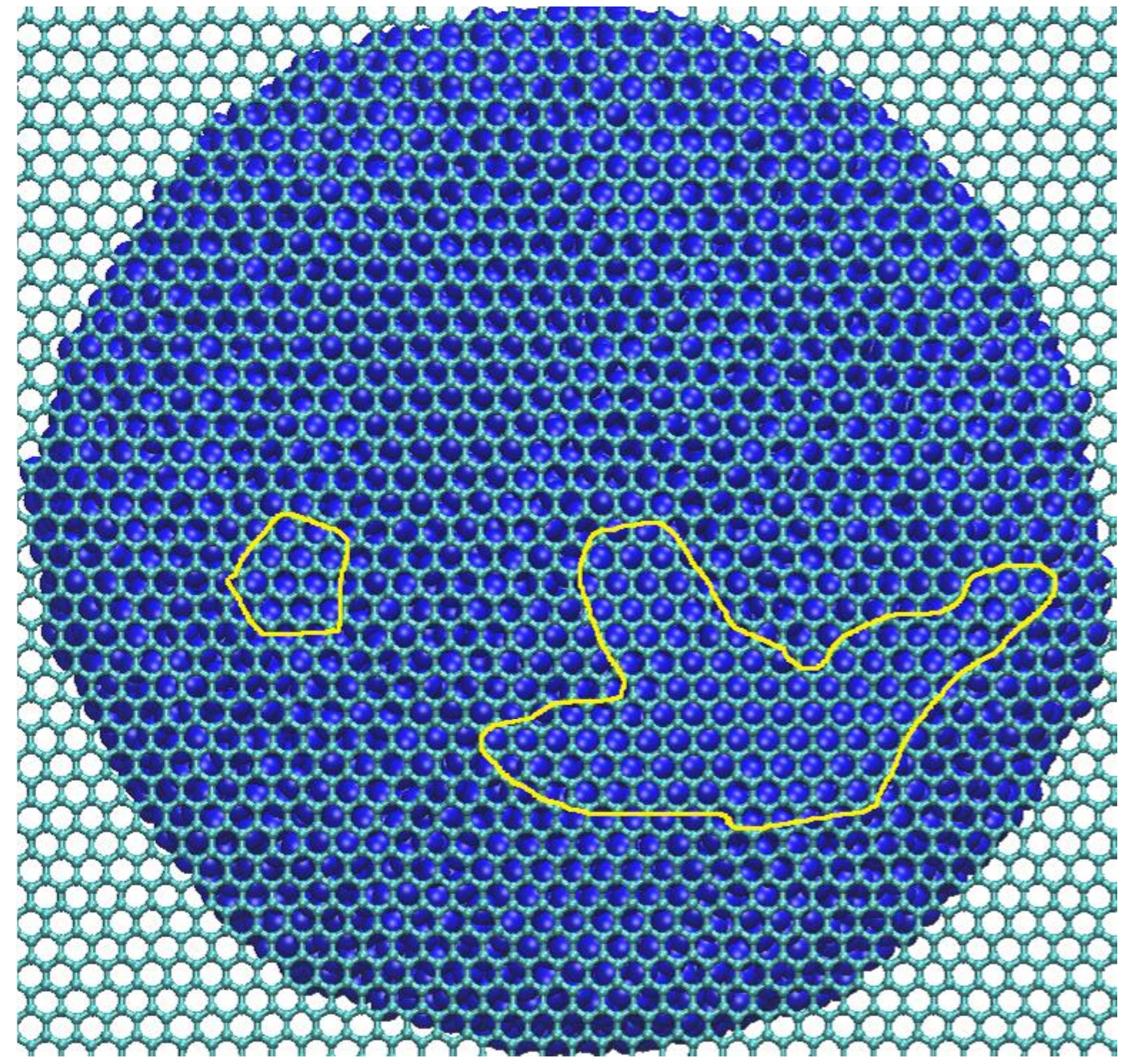}
\includegraphics[width=0.38\textwidth]{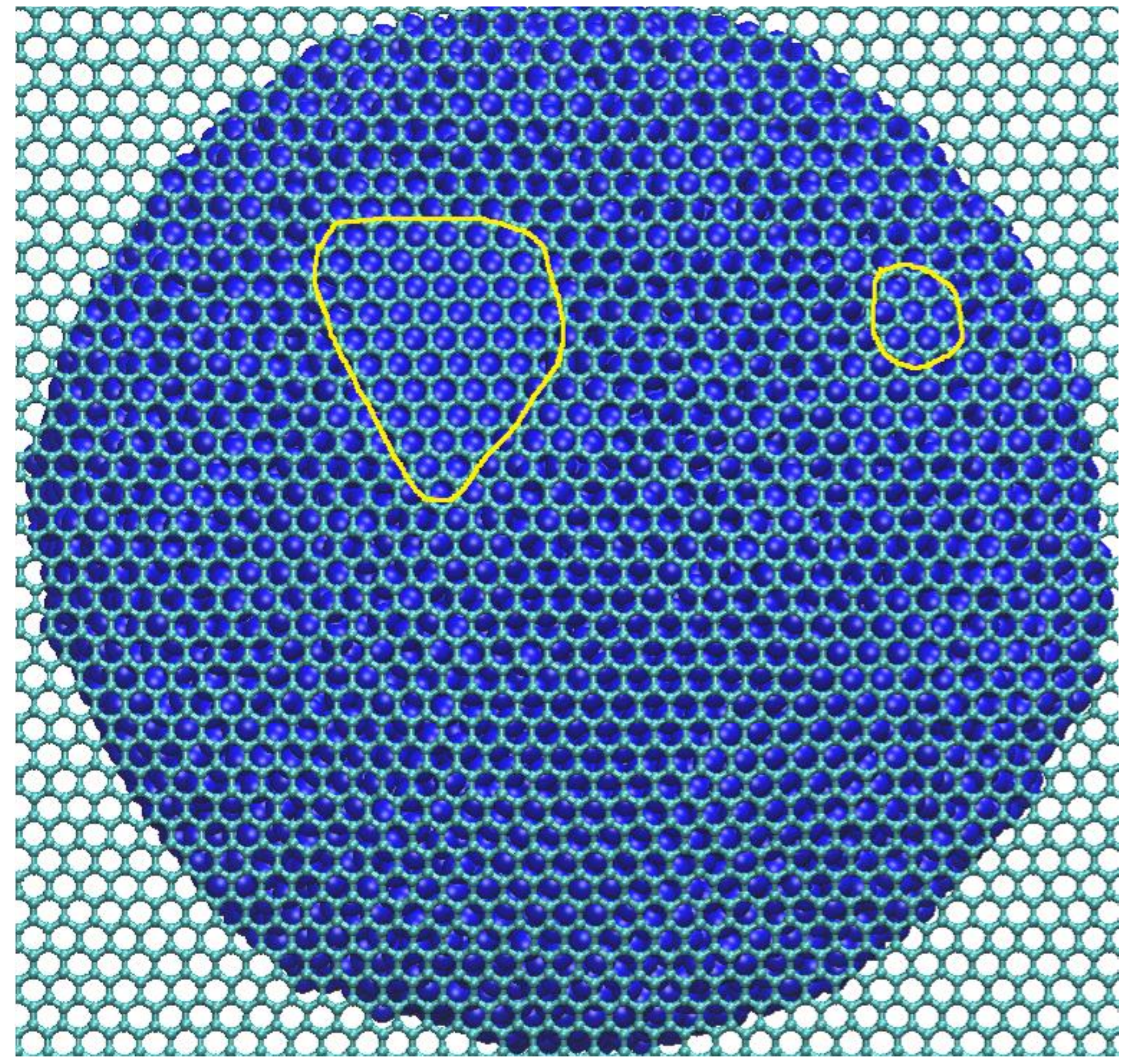}}
\caption{Illustration of contact ``patches'' for Ni NP containing 25000 atoms and sheared at angle 60$^{\circ}$ at time moments of 106 and 132~ps. Patches are outlined by contours.}
\label{spots_Ni_25000}
\end{figure}

A``patch'' model can be employed to qualitatively explain the results. It was introduced in Ref.~\cite{Khomenko2013} to find the reasons for the sawtooth dependencies of the friction force for NPs sliding along the $x$-direction. The key point of the model is the assumption that the friction force is mainly defined by the ordered regions of atoms on the surface of the NP. These ordered domains can form commensurate regions with the substrate (see figure~\ref{spots_Ni_25000}) which give the major contribution to the friction. In order to get high friction and sawtooth substrate force dependencies, the spots should be easily formed and act coherently during the sliding of the NP. This was observed for Ni~\cite{Khomenko2010jpc} and Cu~\cite{Khomenko2013} where the formation of the ``patches'' is facilitated by a good fit of the metal nearest-neighbor distance to the graphene lattice. On the other hand, for Ag and Au the formation of patches is more difficult and hence the sawtooth dependencies were observed only as exception at low sliding velocities. The anisotropy of the friction force can also be attributed to the dependence of the number and the topography of patches on the sliding direction. Those directions where the number of patches is larger are characterized by higher friction. Additionally, in some directions the patches do not act coherently leading to irregular forces, while in other directions the coherent action leads to sawtooth curves. This reasoning is consistent with our simulations, where the force curves are either irregular or sawtooth for Ag depending on the sliding direction of the NP. Quantification of the ``patch'' model and its application to the corresponding systems will be presented in future studies.

\section{Conclusion}\label{conclus}

In conclusion, classical MD simulations have revealed the dependence of the friction force on the shearing direction for Ni and Ag nanoislands. It was shown that the anisotropy effect can lead to the difference in friction of a factor of 2. The dependencies of the components of the substrate force on the corresponding lateral components of the CM coordinates of the NP are also changed with the shear direction. The possibility of the sawtooth dependency was shown for Ag NPs. In this case the frequency of spikes differs from the one of the sawtooth friction curves for Ni. A qualitative ``patch'' model was used as a first effort to explain the friction anisotropy phenomenon. Further advancement of the model to the quantitative level and its application to different systems will be done in future studies.

\section*{Acknowledgements}

A.V.K. thanks Dr. Bo N.J. Persson for invitation, hospitality, helpful comments and suggestions during his stay at the Forschungszentrum J$\rm\ddot u$lich (Germany). A.V.K. is also grateful to Dr.~Persson and the organizers of the conference ``Joint ICTP-FANAS Conference on Trends in Nanotribology'' (12-16 September 2011, Miramare, Trieste, Italy) for invitation and financial support for participation. The work was supported by the grant of the Ministry of Education and Science, Youth and Sports of Ukraine ``Modeling of friction of metal nanoparticles and boundary liquid films which interact with atomically flat surfaces'' (No.~0112U001380) and by the grant for a research visit to the Forschungszentrum J$\rm\ddot u$lich (Germany).


\section*{References}

\begin{thebibliography}{10}
\bibitem{Socoliuc2006} Socoliuc~A, Gnecco~E, Maier~S, Pfeiffer~O, Baratoff~A, Bennewitz~R and Meyer~E 2006 \textit{Science} \textbf{313} 207

\bibitem{Gnecco2007} Gnecco~E and Meyer~E (ed) 2007 \textit{Fundamentals of Friction and Wear on the Nanoscale} (Berlin: Springer)

\bibitem{Bhushan2008} Bhushan~B (ed) 2008 \textit{Nanotribology and Nanomechanics} (Berlin: Springer)

\bibitem{Pantazi2008} Pantazi~A \textit{et al} 2008 \textit{IBM J. Res. Dev.} \textbf{52} 493

\bibitem{Pogrebnjak2009} Pogrebnjak~A~D, Shpak~A~P, Azarenkov~N~A and Beresnev~V~M 2009 \textit{Physics -- Uspekhi} \textbf{52} 29

\bibitem{Pogrebnjak2009vac} Pogrebnjak~A~D \textit{et al} 2009 \textit{Vacuum} \textbf{83} S235

\bibitem{Guerra2010} Guerra~R, Tartaglino~U, Vanossi~A and Tosatti~E 2010 \textit{Nature Materials} \textbf{9} 634

\bibitem{Lyashenko2011} Lyashenko~I~A, Khomenko~A~V and Metlov~L~S 2011 \textit{Tribol. Int.} \textbf{44} 476

\bibitem{Ritter2005} Ritter~C, Heyde~M, Stegemann~B, Rademann~K and Schwarz~U~D 2005 \textit{Phys. Rev.} B \textbf{71} 085405

\bibitem{Dietz2008} Dietzel~D, Ritter~C, M\"{o}nninghoff~T, Fuchs~H, Schirmeisen~A and Schwarz~U~D 2008 \textit{Phys. Rev. Lett.} \textbf{101} 125505

\bibitem{Dietz2009} Dietzel~D, Feldmann~M, Fuchs~H, Schwarz~U~D and Schirmeisen~A 2009 \textit{Appl. Phys. Lett.} \textbf{95} 053104

\bibitem{Schirmeisen2009} Schirmeisen~A and Schwarz~U~D 2009 \textit{ChemPhysChem} \textbf{10} 2373

\bibitem{Dietz2010tl} Dietzel~D, Feldmann~M, Herding~C, Schwarz~U~D and Schirmeisen~A 2010 \textit{Tribol. Lett.} \textbf{39} 273

\bibitem{Dietz2010} Dietzel~D, M\"{o}nninghoff~T, Herding~C, Feldmann~M, Fuchs~H, Stegemann~B, Ritter~C, Schwarz~U~D and Schirmeisen~A 2010 \textit{Phys. Rev.} B \textbf{82} 035401

\bibitem{Aruliah2005} Aruliah~D~A, M\"{u}ser~M~H and Schwarz~U~D 2005 \textit{Phys. Rev.} B \textbf{71} 085406

\bibitem{Brndiar2011} Brndiar~J, Turansk\'{y}~R, Dietzel~D, Schirmeisen~A and Stich~I 2011 \textit{Nanotechnology} \textbf{22} 085704

\bibitem{Khomenko2010jpc} Khomenko~A~V and Prodanov~N~V 2010 \textit{J. Phys. Chem.} C \textbf{114} 19958

\bibitem{Khomenko2013} Khomenko~A~V, Prodanov~N~V, Persson~B~N~J 2013 \textit{arXiv:cond-mat/1302.6432}.

\bibitem{Baur1998} Baur~C, Bugacov~A, Koel~B~E, Madhukar~A, Montoya~N, Ramachandran~T~R, Requicha~A~A~G, Resch~R and Will~P 1998 \textit{Nanotechnology} \textbf{9} 360

\bibitem{Paolicelli2009} Paolicelli~G, Rovatti~M, Vanossi~A and Valeri~S 2009 \textit{Appl. Phys. Lett.} \textbf{95} 143121

\bibitem{Rovatti2011} Rovatti~M, Paolicelli~G, Vanossi~A and Valeri~S 2011 \textit{Meccanica} \textbf{46} 597

\bibitem{Filleter2009} Filleter~T, McChesney~J~L, Bostwick~A, Rotenberg~E, Emtsev~K~V, Seyller~Th, Horn~K and Bennewitz~R 2009 \textit{Phys. Rev. Lett.} \textbf{102} 86102

\bibitem{Lee2009} Lee~C, Wei~X, Li~Q, Carpick~R, Kysar~J~W and Hone~J 2009 \textit{Phys. Stat. Sol.} B \textbf{246} 2562

\bibitem{Lee2010} Lee~C, Li~Q, Kalb~W, Liu~X-Z, Berger~H, Carpick~R~W and Hone~J 2010 \textit{Science} \textbf{328} 76

\bibitem{Wijn2011} de Wijn~A~S, Fasolino~A, Filippov~A~E, Urbakh~M 2011 \textit{Europhysics Letters} \textbf{95} 66002

\bibitem{JNEPH2009} Khomenko~O~V, Prodanov~M~V and Scherbak~Yu~V 2009 \textit{J. Nano- Electron. Phys.}
\textbf{1} No.2 66

\bibitem{Abargues2009} Abargues~R, Gradess~R and Canet-Ferrer~J 2009 \textit{New Journal of Chemistry} \textbf{33} 913

\bibitem{Jeon2010} Jeon~S-H, Xu~P, Mack~N~H, Chiang~L~Y, Brown~L and Wang~H-L 2010 \textit{J. Phys. Chem.} C \textbf{114} 36

\bibitem{Geissler2010} Geissler~A, He~M, Benoit~J-M and Petit~P 2010 \textit{J. Phys. Chem.} C \textbf{114} 89

\bibitem{Nethravathi2011} Nethravathi~C, Anumol~E~A, Rajamathi~M and Ravishankar~N 2011 \textit{Nanoscale} \textbf{3} 569

\bibitem{Park2005} Park~J~Y, Ogletree~D~F, Salmeron~M, Ribeiro~R~A, Canfield~P~C, Jenks~C~J and Thiel~P~A 2005 \textit{Science} \textbf{309} 1354

\bibitem{Filippov2010} Filippov~A~E, Vanossi~A and Urbakh~M 2010 \textit{Phys. Rev. Lett.} \textbf{104} 074302

\bibitem{Gnecco2010} Gnecco~E 2010 \textit{Europhys. Lett.} \textbf{91} 66008

\bibitem{Braun2011} Braun~O~M and Manini~N 2011 \textit{Phys. Rev.} E \textbf{83} 021601

\bibitem{Marcus2002} Marcus M S, Carpick R W, Sasaki D Y and Eriksson M A 2002 \textit{Phys. Rev. Lett.} \textbf{88} 226103

\bibitem{Khomenko2010carbon} Khomenko~A~V and Prodanov~N~V 2010 \textit{Carbon} \textbf{48} 1234

\bibitem{JNEPH2011} Khomenko~A~V and Prodanov~N~V 2011 \textit{J. Nano- Electron. Phys.} \textbf{3} No.2 34

\bibitem{Humphrey1996} Humphrey~W, Dalke~A and Schulten~K 1996 \textit{J. Molec. Graphics.} \textbf{14} 33
    %http://www.ks.uiuc.edu/Research/vmd/.

\bibitem{Sasaki1996} Sasaki~N, Kobayashi~K and Tsukada~M 1996 \textit{Phys. Rev.} B \textbf{54} 2138

\bibitem{Zhou2001} Zhou~X~W \textit{et al} 2001 \textit{Acta Mater.} \textbf{49} 4005

\bibitem{Rapaport2004} Rapaport~D~C 2004 \textit{The Art of Molecular Dynamics Simulation} (Cambridge: Cambridge University Press)

\bibitem{Griebel2007} Griebel~M, Knapek~S and Zumbusch~G 2007 \textit{Numerical Simulation in Molecular Dynamics} (Berlin: Springer)
     %http://wissrech.ins.uni-bonn.de/research/projects/tremolo/download.html

\bibitem{Neek2009} Neek-Amal~M, Asgari~R and Rahimi Tabar~M~R 2009 \textit{Nanotechnology} \textbf{20} 135602

\bibitem{nap2012} Khomenko~A~V, Prodanov~N~V and Sinko~D~O 2012 \textit{Proceedings of the International Conference Nanomaterials: Applications and Properties, 2012} (Sumy: Sumy State University) \textbf{1} No.1 01PCN18(4pp)

\bibitem{Pers_2006} Tartaglino~U, Samoilov~V~N and Persson~B~N~J 2006 \textit{J. Phys.: Condens. Matter} \textbf{18} 4143

\end{thebibliography}
\end{document}